# Multiple change-point Poisson model for threshold exceedances of air pollution concentrations

János Gyarmati-Szabó [a,b,*], Leonid V. Bogachev [b], Haibo Chen [a]

[a] *Institute for Transport Studies, University of Leeds, Leeds LS2 9JT, United Kingdom*
[b] *Department of Statistics, University of Leeds, Leeds LS2 9JT, United Kingdom*

A B S T R A C T

A Bayesian multiple change-point model is proposed to analyse violations of air quality standards by pollutants such as nitrogen oxides ($NO_2$ and $NO$) and carbon monoxide ($CO$). The model is built on the assumption that the occurrence of threshold exceedances may be described by a non-homogeneous Poisson process with a step rate function. Unlike earlier approaches, our model is not restricted by a predetermined number of change-points, nor does it involve any covariates. Possible short-range correlations in the exceedance data (e.g., due to chemical and meteorological factors) are removed via declusterisation. The unknown rate function is estimated using a reversible jump MCMC sampling algorithm adapted from Green (1995), which allows for transitions between parameter subspaces of varying dimension. This technique is applied to the 17-year (1993–2009) daily $NO_2$, $NO$ and $CO$ concentration data in the City of Leeds, UK. The results are validated by running the MCMC estimator on simulated data replicated via a posterior estimate of the rate function. The findings are interpreted and discussed in relation to some known traffic control actions. The proposed methodology may be useful in the air quality management context by providing quantitative objective means to measure the efficacy of pollution control programmes.

*Keywords:* Air quality action plan; Urban air pollution; Threshold exceedances; Non-homogeneous Poisson process; Multiple change-points; Reversible jump MCMC

## 1. Introduction

*1.1. Air quality standards*

Poor air quality, especially in urban areas, is a matter of growing worldwide concern among scientists, policy makers and the general public alike, due to harmful effects of pollution on the human health and the environment. It is well documented that significant health problems (e.g., respiratory and cardiovascular) can be caused or worsened by exposure to the air pollution both on a short- and long-term basis, with the level of severity varying from relatively mild effects, such as an increased use of inhalers by asthmatics, to most serious ones including advancement of the death of individuals (Brimblecombe, 1998; Ayres, 2002).

The UK National Air Quality Strategy (AQS), adopted in 1997 and last reviewed in 2007 (see AQS, 2007), aims at delivering a higher degree of protection against air pollution risks to public health in the UK. It has determined statutory standards for eight main air pollutants, including nitrogen dioxide ($NO_2$), nitrogen monoxide (NO) and carbon monoxide (CO), with tight time frames to achieve the objectives. For example, the $NO_2$ standard (set to be achieved by 31 December 2005) has specified that the hourly mean concentration should not exceed the threshold of 200 µg m$^{-3}$ more than 18 times in one year. Since December 1997, the local authorities are also required to identify Air Quality Management Areas (AQMA) where appropriate (see AQMA, 1997), by setting out an Air Quality Action Plan in pursuit of the air quality objectives.

---

[*] Corresponding author. Tel.: +44 (0)113 343 1788; fax: +44 (0)113 343 5334.
  *E-mail address:* tsjs@leeds.ac.uk (János Gyarmati-Szabó).

It is well documented that the pollutant emission to the atmosphere is dominated by road transport, estimated to contribute 54% of CO and 32% of $NO_x$ emissions in the UK (NAEI, 2010). Moreover, in some zones close to roadside, road traffic can be responsible for more than 90% of the total emission of harmful pollutants such as $NO_2$ (AQAP, 2004, p. 6). The UK Traffic Management Act 2004 (TMA, 2004, with several amendments adopted in 2004–2009) has put in place legislation, implemented from 1 April 2008, aiming to promote a systematic integration of air quality management with transport planning. As a result of these measures, including tough regulations for industry and tightened emission standards for vehicles (EUR, 2003), a much better air quality has been achieved in the UK in general, even though certain elevated levels of pollution may still occur (AQS, 2007; Faulkner and Russell, 2010).

Leeds City Council is among the UK local authorities that are taking a proactive stance in monitoring and management of air quality by deploying the state-of-the-art instrumentations in the city (with the data accessible to transportation and environmental researchers and modellers), and by implementing traffic control schemes. In particular, two AQMAs were declared in small areas of the inner city in Leeds (AQAP, 2004), with the corresponding action plans including measures such as traffic demand management (to reduce the need to travel), improvements to the highways network (to reduce vehicle emissions), reducing emissions from industrial and domestic sources, and raising awareness.

In order to monitor and assess the efficacy of these and future policies, it is important to develop adequate statistical methods and techniques enabling one to measure the impact of the regulations (if any) on the spatiotemporal dynamics of various pollutants, especially with regard to the set standards. It is evident that any such methods should be linked to the data about the exceedance episodes, when the concentrations of one or more pollutants violate certain thresholds. This underpins the importance of developing reliable methods for statistical modelling of exceedances, including their prediction and control. Progress in this direction would provide the necessary feedback helping to assess and improve the current environmental policies, thus avoiding a grossly inefficient and costly "trial and error" approach. Furthermore, in addition to the environmental and health benefits, advancement of research methodologies in the statistical modelling of pollutant concentration extremes could have a significant economic value, potentially leading to massive savings in the public spending related to the government's environmental policies.

*1.2. Modelling of extreme pollution concentration levels*

To date, a number of statistical methods have been developed to model violations of the air quality standards. The common approach is to describe threshold exceedances using extreme value theory, which dates back to Singpurwalla (1972), Horowitz and Barakat (1979), Berger et al. (1982), Surman et al. (1987), Smith (1989) and, more recently, Kütchenhoff and Thamerus (1996), Sharma et al. (1999), Lu and Fang (2003) and Eastoe and Tawn (2009); see also the review papers by Roberts (1979a, 1979b), Thompson et al. (2001), Smith (2004) and further references therein. Other methods have also been proposed, such as Bernoulli, Poisson and mixed-Poisson models (Javits, 1980; Raftery, 1989). However, most of these approaches assume temporal homogeneity of stochastic processes involved (in particular, with a constant exceedance rate), which is hardly valid in the context of non-stationary pollution concentrations, especially over longer time horizons.

To address this issue, some authors have modelled exceedances of air pollution concentrations using non-homogeneous Poisson processes (see, e.g., Raftery, 1989; Smith, 1989; Shively, 1991; Smith and Shively, 1995; Achcar et al., 2008). The idea behind this approach stems from the well-developed asymptotic theory of stationary point processes stating that the series of occurrences of high threshold exceedances can be approximated with a (homogeneous) Poisson process (see, e.g., Leadbetter, 1972; Leadbetter et al., 1983). For nonstationary data, the approximating Poisson process has to be non-homogeneous, with the intensity rate depending on time. Estimation of the unknown rate function is greatly facilitated by choosing certain parametric classes of functions. For instance, Achcar et al. (2008) have used this approach to model violations of the air quality standard in daily ozone measurements in Mexico City using an exponentiated Weibull form for the rate function (Muldholkar et al., 1995). However, such models are not suitable if the rate function may change abruptly (even discontinuously), for instance due to an updated action plan introduced by the local or central authorities in their effort to reduce emission. To capture this type of variation in the exceedance dynamics, Achcar et al. (2009a) proposed a non-homogeneous Poisson model with a single change-point in the rate function, followed by Achcar et al. (2009b) who developed an improved method to handle up to three change-points, based on iterative updates of the prior information.

*1.3. Scope of the paper*

In the present paper, we use a non-homogeneous Poisson model with no prior restriction on the number of change-points. For simplicity, it is assumed that the varying rate $\lambda(t)$ is a step function of time (i.e., it is constant between the neighbouring change-points). However, $\lambda(t)$ is not required to be monotone in time, which was a common restriction in earlier works (see, e.g., Achcar et al., 2009a, and further references therein). The statistics of threshold exceedances will usually involve short-range correlations, because elevated levels of a pollutant will typically occur in patches, perhaps extending over several days, if not weeks. Therefore, after removing an intrinsic seasonal periodicity in the data, it is essential to apply a suitable thin-ning (declusterisation) of the threshold exceedance data, in order to ensure a better fit of the Poisson model (see, e.g., Leadbetter, 1972, Leadbetter et al., 1983, and Shively, 1991; cf. Achcar et al., 2009a, 2009b, where this issue was not addressed). In addition, our approach involves imputation of missing values in the observation data set which could otherwise adversely affect the estimated parameters of the model. An efficient reversible jumps MCMC estimator is then used to automatically determine the number and the time occurrences of change-points, as well as the respective values (heights) of the rate function between the change-points. These techniques are applied to the $NO_2$, NO and CO data collected in the Leeds city centre as daily concentration maxima over nearly 17 years, from 4 January 1993 to 31 December 2009 (AQA, 2010). Our preliminary analysis suggests that some of the identified change-points admit interpretation in the context of the past air quality control actions, but this needs a more thorough investigation.

The paper is organised as follows. In Section 2, we briefly review the general definition and properties of non-homogeneous Poisson processes. Formal tests (both classical and Bayesian) to compare different Poisson models are also described (Section 2.3). Section 3 deals with issues of the model specification and preprocessing, including thresholding, imputing missing values and deseasonalisation of the raw data, as well as declusterisation of threshold exceedances. Bayesian approach to the model estimation is set out in Section 4, including a detailed description of the reversible jump MCMC algorithm. The results of fitting a non-homogeneous (step rate) Poisson model to the $NO_2$, NO and CO threshold exceedance data are reported in Section 5, followed by the model verification and validation. Section 6 contains the discussion, including some preliminary interpretation of results in terms of the past air quality related events and actions, complemented by an outline of directions for future research. Finally, a brief summary of the work is given in Section 7.

## 2. Non-homogeneous Poisson process

*2.1. Definition and basic properties of the Poisson process*

Non-homogeneous Poisson process (see, e.g., Cox and Lewis, 1966) is a statistical model designed to describe a series of uncorrelated random events (in a generic setting usually referred to as "arrivals", motivated by the queue theory). A time-dependent rate function $\lambda(t)$ of the Poisson process determines its local intensity, i.e., the probability of arrival between times $t$ and $t+h$ (with $h \approx 0$) can be approximately expressed in the form $\lambda(t)h + o(h)$, where $o(h)$ indicates terms of higher order of smallness as compared to $h$. This assumption implies that the total number $N(t)$ of arrivals on the interval $[0, t]$ is distributed according to a Poisson law,

$$P\{N(t) = n\} = \frac{\Lambda(t)^n}{n!} e^{-\Lambda(t)}, \qquad n = 0, 1, 2, \ldots, \tag{1}$$

where the cumulative rate function $\Lambda(t) = \int_0^t \lambda(u)\,du$ gives the expected number of arrivals by time $t$. Note that $\Lambda'(t) = \lambda(t)$, so the rate function determines the local expected frequency of arrivals. More generally, the number $N(t) - N(s)$ of arrivals between times $s \geq 0$ and $t > s$ has the Poisson distribution with parameter $\Lambda(t) - \Lambda(s)$.

It follows that the waiting time $\tau$ to first arrival after $s \geq 0$ has the distribution

$$P\{\tau > t - s\} = P\{N(t) - N(s) = 0\} = e^{-\Lambda(t) + \Lambda(s)} \qquad (t > s).$$

Hence, the joint density of $n$ consecutive arrivals restricted to a time interval $[0, T]$ is

$$f(t_1, \ldots, t_n) = \prod_{i=1}^{n} e^{-\Lambda(t_i) + \Lambda(t_{i-1})} \lambda(t_i) \times e^{-\Lambda(T) + \Lambda(t_n)} = e^{-\Lambda(T)} \prod_{i=1}^{n} \lambda(t_i) \qquad (2)$$

$(0 = t_0 < t_1 < \cdots < t_n < T)$,

and the corresponding log-likelihood function is given by

$$\log L(\vec{\theta}; t_1, \ldots, t_n) \equiv \log f(t_1, \ldots, t_n) = \sum_{i=1}^{n} \log \lambda(t_i) - \Lambda(T) \qquad (3)$$

$(0 < t_1 < \cdots < t_n < T)$,

where $\vec{\theta}$ is a (vector) parameter of the model. Formulas (1), (2) imply that the joint density of arrivals in $[0, T]$ conditioned on their total number $N(T) = n$ equals

$$f_n(t_1, \ldots, t_n) = \frac{n!}{\Lambda(T)^n} \prod_{i=1}^{n} \lambda(t_i) \qquad (0 < t_1 < \cdots < t_n < T), \qquad (4)$$

which coincides with the distribution of the order statistics (arranged in the increasing order) for a sample of $n$ independent points thrown into the interval $[0, T]$ with the density $\lambda(t)/\Lambda(T)$ each (e.g., for a constant rate this amounts to the uniform distribution on $[0, T]$). By the usual quantile transformation $U = F(X)$ it follows that the transformed arrivals $t'_i = \Lambda(t_i)/\Lambda(T)$ have the same distribution as the *uniform* order statistics on $[0, T]$ (see Lewis, 1972; Lewis and Shedler, 1976a). Using this fact, one can construct approximate goodness-of-fit tests for particular classes of the rate functions after a suitable estimation of the parameters specifying the model.

Simulation of a non-homogeneous Poisson process may be carried out by the so-called "acceptance/rejection" method based on formula (4) (see Lewis and Shedler, 1976b; Robert and Casella, 2004). Namely, suppose there is a constant $\lambda^*$ such that $\lambda(t) \leq \lambda^*$ for all $t \in [0, T]$. A homogeneous Poisson process with constant rate $\lambda^*$ is simulated and its successive arrival times $t_i$ are either accepted, with probability $\lambda(t_i)/\lambda^*$, or rejected, with the complementary probability $1 - \lambda(t_i)/\lambda^*$; then the accepted sequence corresponds to a Poisson process with the rate function $\lambda(t)$, as required. Note that a Poisson process with $\lambda(t) \equiv \lambda^*$ can be simulated directly using an independent sequence of exponentially distributed inter-arrival times with parameter $\lambda^* T$ (see equation (2)). Alternatively, one can first choose the number of arrivals on $[0,T]$ as a sample value $N(T) = n$ of a Poisson random variable $N(T)$ with parameter $\lambda^* T$ (see equation (1)), and then simulate the corresponding arrival times as an independent sample of size $n$ with the uniform distribution on $[0,T]$ (cf. equation (4)). The direct method can also be used in the case of a step rate function $\lambda(t)$ by simulating random arrivals on each interval where $\lambda(t)$ is constant.

## 2.2. Various classes of the rate functions

A non-homogeneous Poisson process (with an unknown rate function $\lambda(t)$) may be used to model the time series of exceedances (i.e., where a predefined threshold is violated by the observed concentration levels). A number of parametric forms for the rate function $\lambda(t)$ have been used in the air quality modelling, such as the power (Weibull) law $\lambda(t) = \alpha t^{\beta-1}$, Musa–Okumoto's law $\lambda(t) = \beta/(t+\alpha)$, Goel–Okumoto's law $\lambda(t) = \alpha \exp(-\beta t)$ ($\alpha, \beta > 0$), etc. (see Achcar et al., 2008, 2009a, 2009b, and further references therein).

However, these and other conventional formulas are restricted by the monotone (i.e., decreasing or increasing) behaviour of the rate function, which may not be applicable, especially for longer time horizons. To obtain a more flexible form of the rate function, some authors have built models incorporating *covariates* (i.e., hypothetically predictive variables, for example meteorological conditions), in order to make use of any "external" information that may be available (Smith and Shively, 1995).

The approach proposed in the present work does not require the rate function to be monotone, nor does it involve covariates. For simplicity, it is assumed that the rate $\lambda(t)$ is a step function in time,

$$\lambda(t) = \sum_{j=0}^{k} h_j \mathbf{1}\{s_j \leq t < s_{j+1}\}, \qquad 0 = s_0 < s_1 < \cdots < s_k < s_{k+1} = T, \qquad (5)$$

where $\mathbf{1}(A)$ is the indicator of condition $A$ (equal to 1 if $A$ is true and 0 otherwise). Note that the total number $k$ of the change-points, as well as their positions $s_j$ and the heights $h_j$ of the corresponding steps, are not predetermined and therefore have to be statistically estimated.

It should be pointed out that the description of the joint distribution of observed exceedances via a (non-homogeneous) Poisson process is of course not exact, since a continuous-time stochastic process is being used to fit the data available only on a discrete-time (e.g., daily or hourly) grid; for instance, there is at most one exceedance per observation unit. However, as will be demonstrated below (see Section 5), the Poisson-based model with a stepwise rate function (5) furnishes a good modelling approximation for the concentration level data we are studying.

## 2.3. Statistical tests for Poisson models

To assess the utility of the change-point Poisson process (i.e., with a step rate function) as a fitting model for the exceedance data (in our case, with a suspected downward trend suggested by cumulative plots), we can carry out some statistical tests to compare three different Poisson models:

- $M_0$: homogeneous Poisson process (i.e., with a constant rate);
- $M_1$: Poisson process with a decreasing (log-linear) rate, $\lambda(t) = \alpha \exp(-\beta t)$ (also known as Goel–Okumoto's model, see Section 2.2);
- $M_2$: Poisson model with one change-point.

For a threshold exceedance data set $X: t_1 < t_2 < \cdots < t_n$ observed in the interval $[0, T]$, denote

$$S_n = \sum_{i=1}^{n} u_i, \qquad u_i = t_i/T \qquad (i = 1, \ldots, n). \tag{8}$$

A classical goodness-of-fit test for the hypothesis $M_0$ (against any alternative) is based on testing a uniform distribution of arrivals (see Section 2.1) using the statistic

$$U_n = \frac{S_n - n/2}{\sqrt{n/12}}, \tag{9}$$

which has approximately a standard normal distribution under $M_0$ (Cox and Lewis, 1966, p. 47). An alternative (exact) test, sometimes referred to as the *Military Handbook test* (MIL-HDBK, 1981, p. 68; Meeker and Escobar, 1998, p. 409), employs the statistic

$$\chi^2_{2n} = -2 \sum_{i=1}^{n} \log u_i, \tag{10}$$

which has, under $M_0$, a chi-square distribution with $2n$ degrees of freedom. One can also use the general Kolmogorov–Smirnov one-sample test (Cox and Lewis, 1966, Ch. 6; Gibbons and Chakraborti, 2003), based on the comparison of the hypothetical distribution function $F(t)$ (in our case, uniform on $[0, T]$) with the empirical distribution function

$$F_n(t) = n^{-1} \sum_{i=1}^{n} \mathbf{1}\{t_i \leq t\}, \qquad t \in \mathbf{R}.$$

More precisely, the test statistic is defined as

$$D_n = \max_{t \in \mathbf{R}} |F_n(t) - F(t)| \equiv \max\{|u_i - (i-1)/n|, |u_i - i/n|, \ i = 1, \ldots, n\}, \tag{11}$$

its "large" values indicating significant deviations from the null hypothesis $H_0 : F(\cdot)$. This test is useful for nonparametric statistical inference, since the distribution of $D_n$ under $H_0$ does not depend on the function $F(\cdot)$ as long as the latter is continuous.

A test for $M_0$ against $M_2$ may be based on the statistic (Akman and Raftery, 1986b; Raftery, 1989)

$$\Delta_n = \frac{1}{\sqrt{n}} \max\{|g(i-1, u_i)|, |g(i, u_i)|\}, \tag{12}$$

where the maximum is taken over all $u_i = t_i/T$ such that $u_i \in [0.01, 0.99]$, and

$$g(i,u) = i\sqrt{(1-u)/u} - (n-i)\sqrt{u/(1-u)} \qquad (i = 1,\ldots,n, \quad 0 \leq u \leq 1).$$

The asymptotic distribution of $\Delta_n$ as $n \to \infty$ is expressed in terms of the Ornstein–Uhlenbeck process; in particular, the critical values of the test, as well as its p-values, can be evaluated via the approximate formula (see Akman and Raftery, 1986b)

$$P(\Delta_n > z) \approx \sqrt{\frac{2}{\pi}} e^{-z^2/2} \left(\xi z - \xi z^{-1} + z^{-1}\right), \qquad \xi = \log\frac{0.99}{0.01} = 4.595\ldots \tag{13}$$

For the model $M_1$ with the log-linear rate function $\lambda(t) = \alpha \exp(-\beta t)$, the coefficient $\alpha$ is a nuisance parameter which is eliminated by switching to the conditional distribution with $n > 0$ observed arrivals in $[0,T]$. The slope parameter $\beta$ can be estimated by solving (numerically) the maximum likelihood equation $\frac{\partial \log L_n}{\partial \beta}(\hat{\beta}) = 0$, which in view of formula (3) and the log-linear form of $\lambda(t)$ is reduced to (Cox and Lewis, 1966, Ch. 3)

$$\frac{1}{\hat{\beta}T} - \frac{e^{-\hat{\beta}T}}{1-e^{-\hat{\beta}T}} = \frac{S_n}{n} \tag{14}$$

(see the notation (8)). As was explained in Section 2.1, the modified values

$$t'_1 = \frac{\hat{\Lambda}(t_1)}{\hat{\Lambda}(T)}, \quad \ldots, \quad t'_n = \frac{\hat{\Lambda}(t_n)}{\hat{\Lambda}(T)}, \tag{15}$$

based on the estimated cumulative rate $\hat{\Lambda}(t) = \int_0^t \hat{\lambda}(u)\,du = \alpha\hat{\beta}^{-1}\left(1-\exp(-\hat{\beta}t)\right)$, are approximately distributed as the uniform order statistics on $[0,T]$, so that any one of the tests for $M_0$ described above may be applied to assess goodness of fit for $M_1$ (Lewis and Shedler, 1976a).

Models can also be compared using the Bayes factor $B_{ij}$ (Raftery, 1989, 1996), defined as the ratio of posterior to prior odds for $M_i$ against $M_j$:

$$B_{ij} = \frac{P(X|M_i)}{P(X|M_j)}, \qquad X: t_1 < \cdots < t_n. \tag{16}$$

The reference value of the Bayes factor is 1 corresponding to no preference for either of the models, while values greater or smaller than 1 show evidence in favour of the null hypothesis $M_i$ or the alternative hypothesis $M_j$, respectively. It is more convenient to work with the quantity $\beta_{ij} = 2\log B_{ij}$, for which an indicative scale of the strength of evidence for $M_i$ against $M_j$ is given in Table 2 (Raftery, 1996).

**Table 2**

Calibration of the Bayes factor $B_{ij}$ defined in (16) as a measure of evidence to support the null hypothesis $M_i$ against the alternative hypothesis $M_j$.

| $\beta_{ij} = 2\log B_{ij}$ | Evidence for $M_i$ |
|---|---|
| < 0 | Negative (supports $M_j$) |
| 0 to 2 | Barely worth mentioning |
| 2 to 5 | Positive |
| 5 to 10 | Strong |
| > 10 | Very strong |

The Bayes factor for $M_0$ against the log-linear model $M_1$ is given by (Akman and Raftery, 1986a; Raftery, 1989)

$$B_{01} = 0.6449(n-1)\left[\int_0^\infty e^{-S_n y}\left(\frac{y}{1-e^{-y}}\right)^{n-1}dy\right]^{-1}. \tag{17}$$

Furthermore, the Bayes factor for $M_0$ against the change-point model $M_2$ is computed as (Raftery and Akman, 1986; Raftery, 1989)

$$B_{02} = \frac{4\sqrt{\pi}\,\Gamma(n+\tfrac{1}{2})}{\sum_{i=0}^n J_i\Gamma(i+\tfrac{1}{2})\Gamma(n-i+\tfrac{1}{2})}, \tag{18}$$

where $\Gamma(\alpha) = \int_0^\infty x^{\alpha-1}e^{-x}dx$ is the gamma-function, $J_i = \int_{u_i}^{u_{i+1}} x^{-(i+1/2)}(1-x)^{-(n-i+1/2)}dx$ and $u_i = t_i/T$ ($i=1,\ldots,n$), $u_0 = 0$, $u_{n+1} = 1$. Using equation (16), one can also calculate the Bayes factor for $M_1$ against $M_2$ from formulas (17) and (18) as the quotient

$$B_{12} = \frac{B_{02}}{B_{01}}. \tag{19}$$

## 3. Model specification and preprocessing

### 3.1. Thresholding

For the study purposes, a specific threshold must be set in order to extract the statistics of exceedances. To this end, it might seem natural to make use of the current air quality standards (e.g., AQS, 2007). However, there are several difficulties resulting from such a choice: (i) the AQS standard is set out in terms of hourly or maximum daily rolling 8-hour averages, while we work with the daily data; (ii) more importantly, the standard involves not only a threshold (e.g., 200 µg m$^{-3}$ for $NO_2$) but also the maximal permitted frequency (no more than 18 times a year for $NO_2$), which causes the "moving window" type dependence in the exceedance data; (iii) the stand-alone threshold concentration value is high enough to produce scarce statistics of the resulting exceedances (e.g., just 0.11% for $NO_2$). Note also that the AQS thresholds are based on epidemiological studies and therefore are not directly relevant to the analysis of the effects of environmental actions.

A simple convenient alternative is to use the quantile thresholding by choosing, say, a 90th empirical percentile as a threshold for the pollutant concentration, leading to the specific threshold values of 96 µg m$^{-3}$, 185 µg m$^{-3}$ and 2.1 mg m$^{-3}$ for $NO_2$, NO and CO, respectively. Cumulative counts of the resulting exceedance data sets for all three pollutants are shown in Fig. 1 (black plots).

### 3.2. Handling the missing values

Let us point out that the concentration level data we are dealing with in the present paper contain a noticeable proportion of missing values; specifically, approximately 6.6%, 5.5% and 9.9% of the $NO_2$, NO and CO data, respectively, are missing over the observed 17-year period (1993–2009). This considerably decreases the threshold exceedance statistics available and therefore may adversely affect the estimation of the unknown rate function $\lambda(t)$.

Unless there is evidence to the contrary, it is reasonable to assume that the missing data are uninformative, that is, availability of the observational result does not depend on the true concentration level. One rather formal way to circumvent the problem of missing values is to assume, following Smith and Shively (1995), that the rate function $\lambda(t)$ vanishes if a measurement on day $t$ is missing. From equation (2) it is easy to see that the likelihood is automatically adjusted to missing values, since $f(t_1, \ldots, t_n) = 0$ whenever at least one $t_i$ is such that $\lambda(t_i) = 0$. Clearly, this approach leads to under-estimation of the unknown rate function $\lambda(t)$ due to ignoring some potential exceedances. More importantly, though, the underlying convention that the rate function may accidentally vanish on some days is not quite satisfactory, because the possibility of threshold exceedance on those days is artificially ruled out. It also contradicts the adopted Poisson model with a step rate function (where "steps" are implicitly assumed to be constant on reasonably long time intervals).

Instead of effectively ignoring the input of missing values as described above, one could try and compensate for the omissions, aiming to improve the estimation of the rate function (Cox, 1981). In this work, we pursue the simplest approach by imputing the missing values using a (two-sided) moving average estimator. We chose the window size of ±65 days for all three pollutants (note that the earliest missing value occurs on day 135 for $NO_2$). In fact, we first impute *raw* concentration values drawn independently according to their estimated (univariate) distribution, and then apply thresholding to determine if they are to be included in the exceedance data set. Even though such interpolation neglects possible short-range correlations in the concentration time series, it may be expected to work well due to a subsequent high-level thresholding (Leadbetter, 1972; Leadbetter et al., 1983). Alternatively, and more simply, one could impute exceedances directly using the estimated occurrence probability (again based on the moving average techniques); however, the former method is preferable if different threshold values are to be used for the same raw data set.

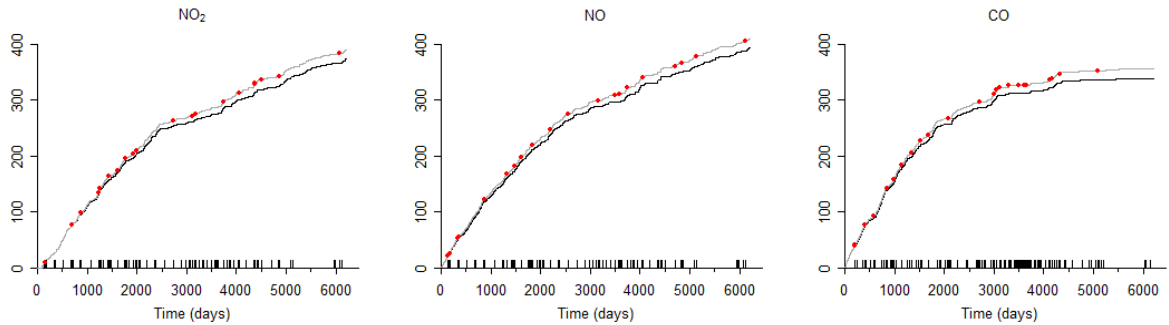

**Fig. 1.** Cumulative count plots of threshold exceedances for the observed raw data (*black*) and for the data with the missing values estimated via the moving-average with a two-sided window of size ±65 days (*grey*). Black ticks on the time axis indicate the occurrence of missing values; red points on the grey curve show the imputed exceedance values.

**Table 1**
Comparison of the MCMC estimation results for $NO_2$ before and after imputation of missing values. Here $n$ = number of exceedances in $[0, T]$; $k$ = number of change-points; $s_1$ = location of the (first) change-point; $h_0, h_1$ = step heights of the rate function on the intervals $(0, s_1)$, $(s_1, T)$, respectively.

| $NO_2$ | $n$ | $k$ | $s_1$ | $h_0$ | $h_1$ |
|---|---|---|---|---|---|
| Before | 619 | 1 | 2493 | 0.1110 | 0.0347 |
| After | 655 | 1 | 2490 | 0.1159 | 0.0366 |

The results of "restoring" the missing values for each pollutant are shown in Fig. 1, where the imputed exceedances are indicated as red points on the cumulative plots. As could be expected, the amended cumulative plots shown in grey lie slightly (but noticeably) higher than the original plots (shown in black). More specifically, Table 1 illustrates the impact of the missing value compensation by comparing the MCMC results for nitrogen dioxide NO$_2$ obtained before and after imputation of the missing values. Note that the number of exceedances has increased by about 5.8%, which is consistent with 6.5% missing raw values for NO$_2$ and some depletion of exceedance values above the empirical 10% threshold following declusterisation (see Section 3.4 below). Both before and after imputation, the number of change-points identified by MCMC was found to be $k = 1$, with almost the same locations; however, the step heights of the rate function noticeably increased (by about 5%) as a result of more exceedance events.

*3.3. Deseasonalisation*

It has to be taken into account that the concentration levels of all three pollutants under study (NO$_2$, NO and CO) are affected by the meteorological conditions which may distort the results. Unfortunately, a detailed specification of the influence of the basic meteorological characteristics (such as spatiotemporal measurements of wind, atmospheric pressure, humidity, sunshine, solar radiation, etc.) is exceedingly complex (Thompson et al., 2001). In addition, information about the concurrent meteorological factors (as well as other covariates) is not readily available from the air quality databases (cf. AQA, 2010).

According to our preliminary investigations, the MCMC method applied to the raw (unprocessed) data may "identify" suspiciously short intervals between some of the estimated change-points, in particular for nitrogen oxides NO$_2$ and NO. A closer inspection reveals that such intervals typically include summer periods with very few exceedances. This may be attributed to the atmospheric photochemical reactions involving ozone O$_3$ (see Clapp and Jenkin, 2001),

$$NO + O_3 \rightarrow NO_2 + O_2, \qquad NO_2 + h\nu\,(+O_2) \rightarrow NO + O_3. \tag{6}$$

The inter-conversion loop (6) underpins the dynamic establishment of equilibrium concentrations of NO$_2$ and NO (so-called "photostationary state"). Note that, due to the asymmetry of reactions (6), the equilibrium concentration levels strongly depend on the presence (or otherwise) of sunlight ($h\nu$), in particular leading to decreased levels of nitrogen oxides NO$_2$ and NO (and, correspondingly, increased levels of ozone O$_3$) during the summer due to higher solar radiation. Carbon monoxide CO is relatively nonreactive, but low temperatures in winter contribute to higher CO concentrations due to incomplete combustion in vehicle engines (CCME, 2002). Because of these and other phenomena, there are yearly cycles in the time series of NO$_2$, NO and CO concentrations, creating "false" change-points in the threshold exceedance rates caused by short-term meteorological variations rather than by new regulations or other anthropogenic factors.

The annual oscillations are clearly visible in the data plots of daily concentration levels for all three pollutants (see Fig. 2, top), which was confirmed by computing the autocorrelation at lag 365 days, giving the values 0.17 for NO$_2$ and NO and 0.33 for CO. Furthermore, analysing the estimated spectral density of the observed time series (Venables, 2002, Ch. 4), the yearly cycles were found to be significant.

There are various possibilities to address the seasonality problem. For instance, following Shively (1991) one could discard the "high ozone season" from May to October each year, where the concentration levels of nitrogen oxides $NO_2$ and $NO$ are likely to be low, and investigate only the complementary winter periods (November–April). However, this approach will automatically confine the search to the winter months, while any possible change-points during the summer will be lost.

An alternative solution may be based on various methods of deseasonalisation (i.e., removing the yearly periodicities from the raw time series $X(t)$), for instance by fitting a simple parametric model for the log-transformed data (Lewis, 1972, §5.2):

$$\log \widetilde{X}(t) = \log X(t) - a\cos\omega t - b\sin\omega t - c, \qquad \omega = \frac{2\pi}{365} \text{ (days}^{-1}\text{)}, \qquad (7)$$

where the unknown parameters $a, b, c$ can be estimated using standard linear regression methods. Using the log transform prior to regression is customary for non-negative time series (Cox and Lewis, 1966; Lewis, 1972), when the signal is typically modulated by a deterministic profile while the noise acts multiplicatively. Even though the GLM regression model may not be quite accurate for the concentration data, for instance due to deviations from stationarity over longer time horizons, it is plausible that any such deviations occur on a temporal scale much slower than the annual variability, hence the regression estimates may be expected to give good enough results (cf. Cox and Lewis, 1966, p. 41). In particular, only the "free" coefficient $c$ in equation (7) is likely to be affected, which however has no impact on the subsequent change-point analysis for threshold exceedances.

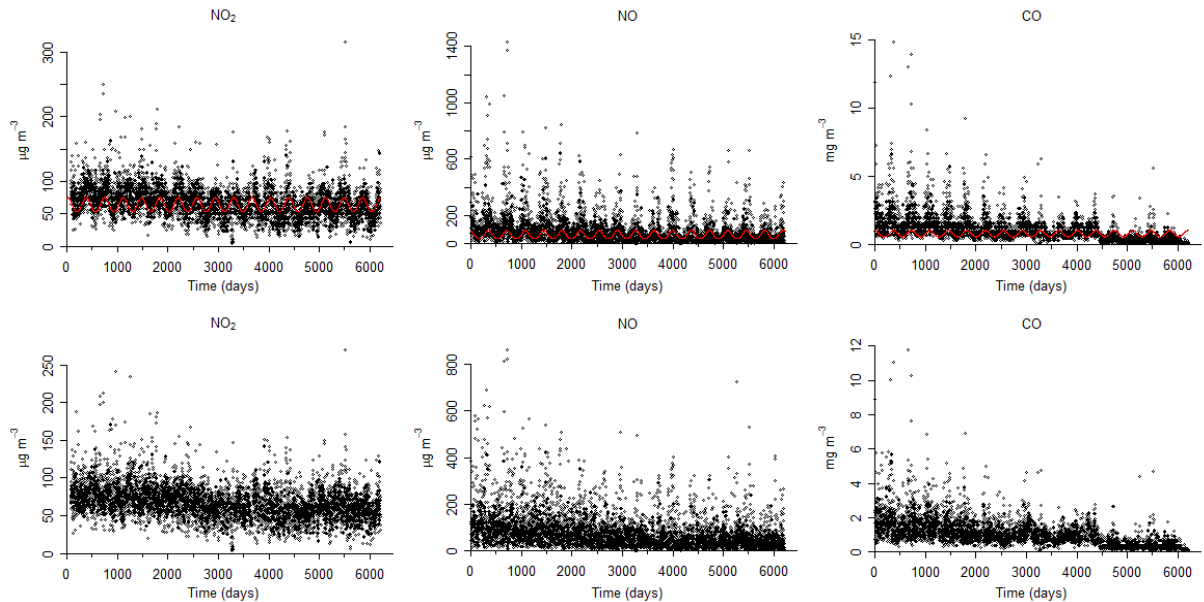

**Fig. 2.** *Top:* observed daily maximum concentrations of $NO_2$, NO and CO (*black*) and the estimated yearly periodic trend for the log-transformed data (*red*). *Bottom:* deseasonalised data, with the yearly periodic trend removed.

The results of deseasonalisation are presented in Fig. 2. Periodic trends filtered out from the raw data using the regression model (7) are shown in Fig. 2 (top) as the red oscillating graphs imposed on top of the raw data plots. Deseasonalised raw data $\widetilde{X}(t)$ are plotted in Fig. 2 (bottom) showing a much improved scattering, with no visible periodic pattern.

Let us point out that alternative methods of deseasonalisation may be used, for example those based on moving-average, kernel type local smoothing or spectral techniques (Cox and Lewis, 1966; Lewis, 1972; Lewis and Shedler, 1976a). One could also take into account the weekly periodicities, which may be anticipated to be present in the data due to traffic patterns caused e.g. by peak hours as well as weekday–weekend oscillations. We have opted not to do so, because the weekly oscillations occur much more frequently as compared to the annual scale, so presumably are better averaged out by fitting a non-homogeneous Poisson model with a step rate function. In addition, the weekly patterns are likely to be more stable across decades, being less affected by any policy changes.

### 3.4. Declusterisation and verification of independence

As was mentioned in Section 2.1, an important characteristic feature of the Poisson model is the so-called "lack of memory", whereby the occurrence of new arrivals after a time instant $t$ is independent of any arrivals that occurred by time $t$. Even though very convenient analytically, this property may be restrictive enough to prevent the immediate applicability of the Poisson modelling approach if the data involve significant correlations. In the air quality context, the dependence in data may manifest itself through clustering (grouping) of the threshold exceedances, since the pollutant concentration having achieved a high value is likely to stay elevated for some time. However, due to a relatively short range of such correlations in the exceedance process, this can be rectified (Leadbetter, 1972; Leadbetter et al., 1983) by applying the Poisson model not to the entire data set of exceedances but to a smaller subset obtained by thinning (also called "declusterisation"), whereby each cluster of exceedances is replaced by a single representing point, say the highest value (i.e., the cluster mode).

There is no universally accepted theoretical method to identify "clusters", so one has to use appropriate empirical rules. Suppose that the data are encoded as a binary sequence of symbols + (plus) and − (minus) representing values, respectively, above and below the threshold. In the present work, the following simple method is adopted: the first plus (+) encountered in the data sequence initiates the first cluster, which remains active until there are $m_0$ consecutive minuses (where $m_0$ is a fixed integer); after the first cluster is defined, the next plus initiates the second cluster, and so on. Recall that the declusterisation procedure keeps only one plus from each cluster, so the corresponding declustered sequence will consist of isolated pluses each followed by at least $m_0$ minuses. Of course, such a sequence cannot be independent, but the independence will be restored if each plus is merged with the next $m_0$ minuses and relabelled as a new plus. In what follows, declusterisation is understood to be carried out together with relabelling just described.

Statistical inference about the significance of correlations in a given data set may be carried out by many methods, for instance using the sample correlation coefficient between adjacent inter-event times (Shively, 1991) or by statistical tests based on estimation of serial correlation at various lags (Solow, 1991). We use the well-known Wald–Wolfowitz runs test (see, e.g., Gibbons and Chakraborti, 2003), based on the assumption that, under the null hypothesis $H_0$ of independence, the (unknown) probabilities of the symbols + and − do not change with time. Denote by $R_n$ the total number of runs in a sample of size $n$, where a *run* is an uninterrupted string of same symbols delimited from both ends by the opposite symbols or by the beginning/end of the sample. Under the hypothesis $H_0$, the distribution of $R_n$ conditioned on the observed number of pluses, $n_+$, is approximately normal with mean and variance

$$\mu_n = 1 + \frac{2n_+(n-n_+)}{n}, \qquad \sigma_n^2 = \frac{(\mu_n-1)(\mu_n-2)}{n-1}.$$

Significant deviations of the observed value $R_n$ from the mean $\mu_n$ (as compared to the standard deviation $\sigma_n$) would suggest that the hypothesis $H_0$ should be rejected.

Note that in the presence of change-points the runs test may strongly reject $H_0$ even though the data may still be independent (e.g., generated by a non-homogeneous Poisson process). Therefore, to make the runs test work one needs to estimate the number and locations of the change-points in advance. The strategy adopted in the present work is as follows.

(i) First, a rough identification of change-points $0 = s_0 < s_1 < ... < s_k < s_{k+1} = T$ is carried out using the MCMC method on the "raw" data set of exceedances (encoded as a +/− sequence).

(ii) The runs test is applied to the raw data on each interval $(s_j, s_{j+1})$, with anticipation that the independence hypothesis $H_0$ will still be rejected due to the correlated clusters of exceedances.

(iii) The data are declustered as explained above, and the runs test is applied once again on each interval $(s_j, s_{j+1})$.

(iv) If the null hypothesis $H_0$ is accepted on all intervals $(s_j, s_{j+1})$, the MCMC identification of the change-points is repeated on the declustered data set, resulting in the adjustment of the rough estimates.

(v) Steps (iii) and (iv) are repeated with the adjusted change-points, in order to ensure that the results are self-consistent.

(vi) Finally, to gain more confidence in the inference obtained, the results can be validated by carrying out the MCMC algorithm on a simulated Poisson data set using the rate function estimated from the data in steps (iv) and (v), and by comparing the observed and simulated counting processes.

We shall report the results of the independence testing via the Wald–Wolfowitz runs test in Section 5.1 below.

## 4. Bayesian reversible jump MCMC parameter estimation

### 4.1. Parametric models of varying dimension

In this section, we describe a Bayesian-based approach that can be used to estimate the unknown step rate function $\lambda(t)$ of a non-homogeneous Poisson process. The method is implemented via a reversible jump Markov chain Monte Carlo (MCMC) sampler based on the Metropolis–Hastings algorithm (see, e.g., Gilks et al., 1996; Brooks, 1998), which is used to compute the posterior distribution of the unknown parameters. The classic MCMC estimator is based on sampling from the posterior distribution, which is proportional, according to the Bayes formula, to the likelihood function and the joint prior distribution of the parameters. In practical terms, the posterior distribution is unavailable analytically, but can be obtained by computer simulations as the equilibrium distribution of a suitable Markov chain in the parameter space, using sufficiently long runs of the chain and taking advantage of its (exponentially fast) convergence to the equilibrium. Algorithmically, the Markov chain is advanced at each step by either accepting or rejecting the proposed "update" values of the parameters.

However, in many modelling situations the dimension of the object of inference is not known in advance, and therefore the classic MCMC method cannot be employed. Our problem is precisely of this kind, since the number of change-points in a step rate function $\lambda(t)$ is not fixed in advance. To circumvent the problem of dimension, Green (1995) has proposed a more flexible MCMC framework by constructing reversible Markov chains allowing for jumps between different parameter subspaces of variable dimension (see Green and Hastie, 2009 and Lunn et al., 2009 for more recent developments). In addition, this methodology resolves the problem of Bayesian model determination, since the MCMC sampler produces the best fitting model. The property of *reversibility*, defined as the detailed balance between possible transitions of the Markov chain, ensures better mixing and faster convergence to equilibrium.

In the present paper, this approach is adapted in the context of extreme value analysis and is applied to threshold exceedance data for concentration levels of several key pollutants (such as $NO_2$, NO and CO).

### 4.2. Prior distribution of the parameters

Let us describe how the prior distribution for the rate function $\lambda(t)$ is chosen in our model. From the cumulative plot of exceedance frequencies for nitrogen dioxide $NO_2$ (Fig. 1), it is evident that the initial large part of the graph, up to $t \approx 2500$, is very close to a linear function, with the slope about $280/2500 = 0.112$, followed by a no-ticeable drop of the slope to approximately $(300 - 280)/(3600 - 2500) = 0.018$ until $t \approx 3600$. One may also suspect one or two less significant change-points further on (e.g., near $t \approx 4500$). Similarly, the cumulative graphs for NO and CO in Fig. 1 suggest about four to five hypothetical change-points each.

Therefore, as a prior distribution for the unknown number $k$ of change-points it is reasonable to use a Poisson probability law,

$$p(k) = \frac{\mu^k e^{-\mu}}{k!} \qquad (k = 0, 1, 2, \ldots), \tag{20}$$

with parameter, say, $\mu = 4.5$. For simplicity (and without loss of generality), we adopt the same prior for each pollutant, even though the cumulative plots for NO and especially $NO_2$ indicate somewhat fewer change-points (see Fig. 1). In fact, as can be anticipated from the general practice of MCMC (Gilks et al., 1996, p. 8) and according to our experimental simulation studies (not reported here), the reversible jump MCMC sampler is not sensitive to a specific choice of the prior distributions for parameters.

To avoid wasteful excursions of the MCMC sampler to excessively large values of $k$, it is convenient to put an upper bound $k \leq k_{max}$ and hence to use the corresponding conditional (truncated) Poisson distribution as a prior for $k$. In this study, the value $k_{max} = 20$ was chosen, which seems to be a realistic bound for the number of environmental actions that might have been introduced during the period of 17 years. Note that the maximum dimension of the parameter space needed for this model is

$$\dim(k) + \dim(s_j) + \dim(h_j) \leq 1 + k_{max} + (k_{max} + 1) = 42. \tag{21}$$

Following Green (1995), for a given $k$ the change-point positions $s_1, \ldots, s_k$ are modelled as even-numbered order statistics amongst $2k + 1$ random points uniformly distributed in $[0, T]$. As compared to a more straightforward sampling of $k$ uniformly distributed points, this approach aims to reduce the occurrence of unrealistically small intervals between the change-points with virtually no data input from there, which would therefore slow down the update dynamics of the MCMC algorithm near such points. The values (heights) $h_j$ of the step function $\lambda(t)$ on the corresponding subintervals $(s_j, s_{j+1})$ ($j = 0, \ldots, k$) are sampled according to exponential distribution with parameter $\gamma = T/N$, where $N = N(T)$ is the total number of observed events (i.e., exceedances) over the observational period $[0, T]$. In particular, the prior expected value of each $h_j$ equals $\gamma^{-1} = N/T$, thus coinciding with the average number of exceedances per day.

## 4.3. Updates of the MCMC sampler states

In this section, we give a brief summary of the algorithmic MCMC updating used in simulations, including proposals for various moves and the corresponding acceptance probabilities; more details can be found in the original paper by Green (1995).

First of all, possible proposals for updates of the MCMC sampler (with the corresponding probabilities in the parentheses indexed by the current number of change-points $k$) are of the following four types:

- changing the height of a randomly chosen step ($\eta_k$);
- changing the position of a randomly chosen change-point ($\pi_k$);
- increasing the number of change-points by one ($b_k$);
- decreasing the number of change-points by one ($d_k$).

Clearly, $\pi_0 = d_0 = 0$, and we also set $b_{20} = 0$ to avoid exceeding the maximum $k_{max} = 20$. For $k > 0$ we take $\eta_k = \pi_k$. The birth and death probabilities $b_k$ and $d_k$ are specified as follows,

$$b_k = C\min\{1, \mu/(k+1)\}, \qquad d_{k+1} = C\min\{1, (k+1)/\mu\} \qquad (k = 0, 1, \ldots, 19), \qquad (22)$$

where a constant $C$ is chosen as large as possible subject to the constraint $b_k + d_k \leq 0.9$. Equation (22) ensures the balance condition $b_k p(k) = d_{k+1} p(k+1)$ (see equation (20)) necessary for the reversibility of birth-death jumps.

If a height change or a move of position is chosen, then the classical fixed-dimensional Metropolis–Hastings sampler is applied (see Gilks et al., 1996). A height change is attempted by first choosing randomly one of the existing heights $h_0, h_1, \ldots, h_k$, say $h_j$, and proposing a new value $h'_j$ such that $\log(h'_j/h_j)$ is uniformly distributed on $[-\frac{1}{2}, \frac{1}{2}]$. The acceptance probability for this move is given by

$$\min\{1, (\text{likelihood ratio}) \times \exp(-\gamma(h'_j - h_j))\},$$

where the *likelihood ratio* is the quotient of the likelihood functions for the proposed and current values of the parameters, respectively (see formula (2)). For a position change, one of the existing change-points $s_1, s_2, \ldots, s_k$ is chosen randomly, say $s_j$. The proposed value $s'_j$ is then drawn uniformly from $[s_{j-1}, s_{j+1}]$, with the corresponding acceptance probability

$$\min\left\{1, (\text{likelihood ratio}) \times \frac{(s_{j+1} - s'_j)(s'_j - s_{j-1})}{(s_{j+1} - s_j)(s_j - s_{j-1})}\right\}.$$

The birth step involves a trans-dimensional move of the chain. First, the position of a new change-point $s^*$ is chosen with a uniform distribution over $[0, T]$, falling in one of the existing intervals, say $(s_j, s_{j+1})$. If this proposal is accepted then a new change-point will be set to $s'_{j+1} = s^*$, while the former change-points $s_{j+1},\ldots,s_k$ will be relabelled as $s'_{j+2},\ldots,s'_{k+1}$, with the corresponding changes in the labels of step heights. The proposed new heights $h'_j$ and $h'_{j+1}$ on subintervals $(s_j, s^*)$ and $(s^*, s_{j+1})$, respectively, are chosen to satisfy the weighted geometric mean condition

$$(s^* - s_j)\log h'_j + (s_{j+1} - s^*)\log h'_{j+1} = (s_{j+1} - s_j)\log h_j \qquad (23)$$

together with the relation

$$\frac{h'_{j+1}}{h'_j} = \frac{1-U}{U},$$

where $U$ is a random value drawn uniformly from $[0,1]$. Detailed balance with the corresponding death move is achieved by reversing the above calculation; namely, if a former change-point $s_{j+1}$ is removed then the new height $h'_j$ over the interval $(s'_j, s'_{j+1}) = (s_j, s_{j+2})$ is chosen subject to the relation (cf. equation (23))

$$(s_{j+1} - s_j)\log h_j + (s_{j+2} - s_{j+1})\log h_{j+1} = (s'_{j+1} - s'_j)\log h'_j.$$

The acceptance probability of the birth proposal is given by

$$\min\{1, (\text{likelihood ratio}) \times (\text{prior ratio}) \times (\text{proposal ratio}) \times (\text{Jacobian})\},$$

where the quantities involved are specified as follows:

$$\text{prior ratio} = \frac{\mu}{k+1} \times \frac{2(k+1)(2k+3)}{T^2} \times \frac{(s^*-s_j)(s_{j+1}-s^*)}{s_{j+1}-s_j} \times \gamma\exp\{-\gamma(h'_j + h'_{j+1} - h_j)\},$$

$$\text{proposal ratio} = \frac{d_{k+1}T}{b_k(k+1)}, \qquad \text{Jacobian} = \frac{(h'_j + h'_{j+1})^2}{h_j}.$$

The acceptance probability for the corresponding death step (i.e., decreasing the number of change-points by one) has the same form with an appropriate relabelling, whereby the ratio terms have to be inverted (Green, 1995).

## 5. Results for the air pollution data

In this section, we report the results obtained by applying the methods described in Sections 3 and 4 to the $NO_2$, NO and CO concentration level measurements collected in the Leeds city centre (at Queen Square Court) by a permanent monitoring station included in the UK automatic urban and rural network (AURN). The station is located about 30 metres from a busy 4-lane inner-city road and some 150 metres from an urban motorway, with traffic flow of approximately 21,500 and 93,500 vehicles per day, respectively (AURN, 2010). The data used in the study (available from AQA, 2010) correspond to the daily maxima for each of the three pollutants observed from 4 January 1993 to 31 December 2009 (i.e., for a total of $T = 6\,206$ days).

## 5.1. Results of declusterisation

Following Section 3.1, the quantile thresholding was used to extract the statistics of exceedances from the original data set of daily pollutant concentrations, with the 90th empirical percentile as a threshold. This was done after imputing the missing values (see Section 3.2), deseasonalisation by removing annual oscillations (Section 3.3) and declusterisation of the raw threshold exceedance data (see Section 3.4).

As was explained in Section 3.4, declusterisation (thinning) is needed in order to eliminate possible correlations in the raw sequence of threshold exceedances, thus making the Poisson model more suitable. Clearly, larger values of the cluster gap $m_0$ would provide more independence, but in practice a specific value of $m_0$ should be chosen as small as possible in order to minimise the loss of information caused by the thinning. In this study, we started with the smallest possible value $m_0 = 1$, motivated by a plausible assumption that daily exceedances one day or more apart may already be treated as independent. Specifically, declusterisation with $m_0 = 1$ resulted in a significant reduction of the data set size by about 30–35% (e.g., for a 90% threshold the data set for $NO_2$ consisted of 612 exceedances with 403 clusters, while for the 95% threshold, there were 311 exceedances with 219 clusters).

The results of declusterisation were quite satisfactory, as was confirmed by the runs test of independence outlined in Section 3.4. To start with, we found that the runs test strongly rejected the hypothesis of independence for the entire data set of exceedances, with extremely small p-values: $2.97 \times 10^{-7}$, $3.13 \times 10^{-6}$ and $8.93 \times 10^{-6}$ for $NO_2$, NO and CO, respectively. This does not necessarily disprove independence, since high significance of the test results could be caused by noticeable non-homogeneities in the data due to the change-points. However, the runs test also rejected the null hypothesis for subsequences between the estimated change-points when applied to the raw series of exceedances (i.e., before declusterisation). On the other hand, the runs test accepted independence for the declustered series on all intervals between the estimated change-points, with the p-values at least 0.11; however, the null hypothesis was still rejected — not surprisingly — for the entire declustered series.

## 5.2. Posterior estimates of the rate function

Let us now describe the results of running the MCMC sampling algorithm (specified in Sections 4.2–4.3) applied to the threshold exceedance data. The programming code for reversible jump MCMC was implemented in R and was run on a standard laptop (AMD processor DualCore 2.2 GB, RAM 4GB). The sampler takes 1000 steps within 20 seconds. The manual programming "from scratch" was deemed necessary since the standard MCMC automatic engines (e.g., WinBUGS, http://www.mrc-bsu.cam.ac.uk/bugs/) are not suitable for models with unspecified parameter dimen-sion (only recently, a trial version for the reversible jump MCMC was made available within WinBUGS, see Lunn et al., 2009).

After a burn-in period of length 15,000 (for $NO_2$) to 20,000 (for NO and CO), pilot runs established that the dynamics of the Markov chain have reached the stationary regime (Gilks et al., 1996; El Adlouni et al., 2006). As usual, convergence of the sampling algorithm was monitored by visual inspection of the output plots and also by simple convergence diagnostic tools based on the MCMC output for a single Markov chain (see a review and further references in Gilks et al., 1996). In addition, reliability of the burn-in runs was verified by trying different initial values, which was also utilised to control the termination time of the MCMC algorithm.

In the stationary regime, the MCMC simulation was run for another 500,000 updates. Examining the corresponding autocorrelation plots, it was determined that in order to ensure independence within the sample, approximately every 40th generated value should be selected and kept for future inference, thus providing a usable posterior sample of size $500{,}000/40 = 12{,}500$ for each pollutant. In view of the parametric dimensional bound given by equation (21), this means that there are at least $12{,}500/42 \approx 300$ sample values per one parameter. In fact, in the estimated equilibrium regime, with $k \leq 4$ (see below), the actual parametric dimension reduces to no more than 10, hence there will be at least $12{,}500/10 = 1250$ values per one parameter, which ensures good accuracy of the posterior estimation.

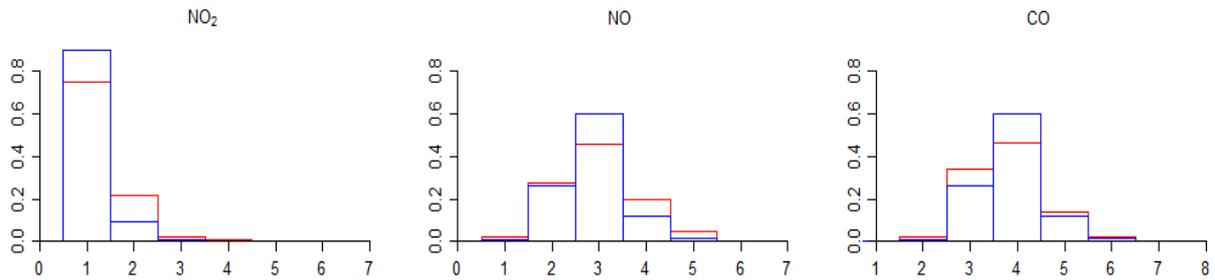

**Fig. 3.** Estimated MCMC posterior distribution of the number $k$ of change-points in the step rate function $\lambda(t)$, obtained for the observed threshold exceedance data (*red*) and for data simulated via non-homogeneous Poisson process with the estimated step function $\hat{\lambda}(t)$ as the rate (*blue*).

As a result of the MCMC performance, it was concluded that the MCMC sampler spent most of its time in parametric states corresponding to $k=1$, $k=3$ and $k=4$ change-points for $NO_2$, NO and CO, respectively (see Fig. 3). The change-point locations $s_j$, estimated via the modes of the respective posterior marginal densities, are shown in Fig. 4. Specific time values of the change-points, along with (quite narrow) 50% credible intervals, are given in Table 3, which should also assist in translation between the times measured in consecutive days (from 1 to 6206) and the corresponding calendar dates (in the format dd-mm-yy). The step heights $h_j$ of the rate function $\lambda(t)$ were estimated via the median of the posterior marginal distributions (see Fig. 5 and Table 4).

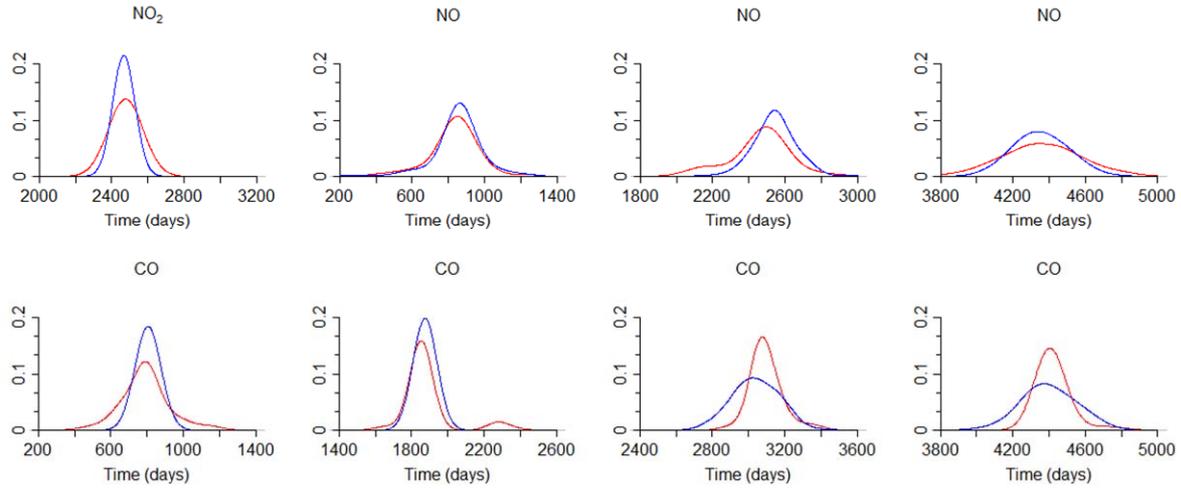

**Fig. 4.** Estimated MCMC posterior density functions of the marginal distributions for the locations $s_i$ of the identified change-points (cf. Table 3), obtained for the observed threshold exceedance data (*red*) and for data simulated via non-homogeneous Poisson process with the estimated step function $\hat{\lambda}(t)$ as the rate (*blue*). The estimates were computed using a Gaussian kernel with bandwidth 95 days.

**Table 3**
Posterior modes of the estimated change-points $s_j$ and the corresponding lower and upper quartiles, expressed as day numbers (ranging from 1 to 6206) as well as calendar dates (dd-mm-yy).

| Pollutant | $NO_2$ | NO | | | CO | | | |
|---|---|---|---|---|---|---|---|---|
| Change-point | $s_1$ | $s_1$ | $s_2$ | $s_3$ | $s_1$ | $s_2$ | $s_3$ | $s_4$ |
| 25th percentile | 2425 25-08-99 | 789 03-03-95 | 2433 02-09-99 | 4281 23-09-04 | 794 08-03-95 | 1788 26-11-97 | 3003 25-03-01 | 4346 27-11-04 |
| Mode | 2490 29-10-99 | 865 18-05-95 | 2503 11-11-99 | 4350 01-12-04 | 822 05-04-95 | 1867 13-02-98 | 3063 24-05-01 | 4388 08-01-05 |
| 75th percentile | 2589 05-02-00 | 912 04-07-95 | 2612 28-02-00 | 4564 03-07-05 | 901 23-06-95 | 1939 26-04-98 | 3174 12-09-01 | 4471 01-04-05 |

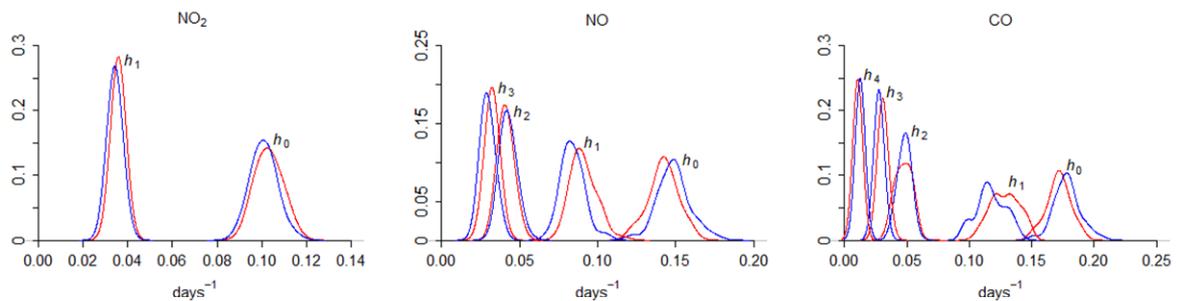

**Fig. 5.** Estimated MCMC posterior density functions of the univariate marginal distributions for the heights $h_j$ of the estimated change-points (cf. Table 4), obtained for the observed threshold exceedance data (*red*) and for data simulated via the estimated step rate function $\hat{\lambda}(t)$ (*blue*). The estimates were computed using a Gaussian kernel with bandwidth 0.003 days$^{-1}$.

**Table 4**
Posterior modes of the estimated heights $h_j$ of the unknown step rate function $\lambda(t)$, with the corresponding lower and upper quartiles.

| Pollutant | NO$_2$ | | NO | | | | CO | | | | |
|---|---|---|---|---|---|---|---|---|---|---|---|
| Height | $h_0$ | $h_1$ | $h_0$ | $h_1$ | $h_2$ | $h_3$ | $h_0$ | $h_1$ | $h_2$ | $h_3$ | $h_4$ |
| 25th percentile | 0.1001 | 0.0321 | 0.1364 | 0.0854 | 0.0352 | 0.0260 | 0.1596 | 0.1095 | 0.0417 | 0.0206 | 0.0021 |
| Mode | 0.1032 | 0.0357 | 0.1421 | 0.0910 | 0.0411 | 0.0334 | 0.1716 | 0.1174 | 0.0473 | 0.0228 | 0.0027 |
| 75th percentile | 0.1071 | 0.0376 | 0.1490 | 0.0978 | 0.0466 | 0.0381 | 0.1831 | 0.1247 | 0.0532 | 0.0247 | 0.0032 |

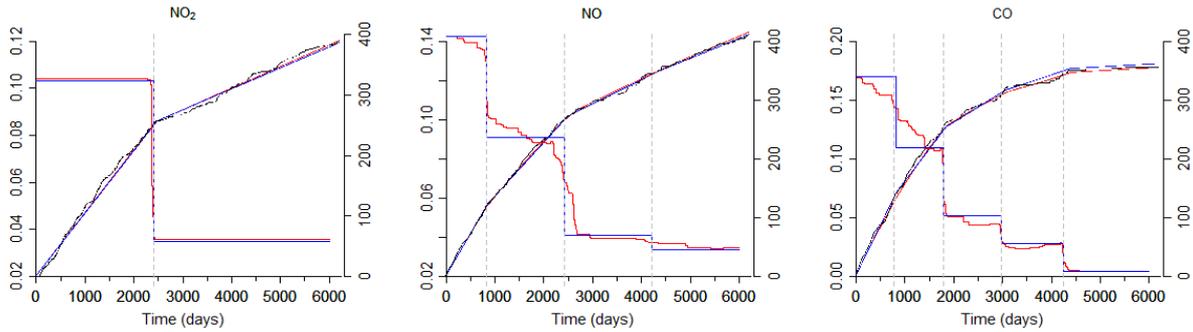

**Fig. 6.** MCMC results for each pollutant: estimated rate functions (*left axes*) and the corresponding cumulative plots for data sets simulated via non-homogeneous Poisson process (*right axes*). Colour-coded plots correspond to the posterior mean rate $\overline{\lambda}(t)$ (*red*) and a step rate function $\hat{\lambda}(t)$ estimated from the posterior distribution (*blue*). *Black* dotted graphs show the cumulative plots of observed threshold exceedances. Dashed vertical lines indicate the locations of estimated change-points.

The corresponding step function $\hat{\lambda}(t)$ (with estimated change-points $s_j$ and step heights $h_j$ as above) may be viewed as a "point" estimate of the unknown rate function $\lambda(t)$ picked from the model class of step functions. The graph of $\hat{\lambda}(t)$ is shown in Fig. 6 (blue, left axes) in comparison with an "integral" estimate $\overline{\lambda}(t)$ (red) of the posterior mean rate function $E(\lambda(t)\,|\,t_1, t_2, \ldots, t_n)$ calculated by averaging 5000 sample step rate functions drawn from the MCMC posterior distribution. Both graphs are fairly close to each other, especially for NO$_2$ and NO but with a more noticeable deviation for CO over the first two steps. In particular, the transition areas for $\overline{\lambda}(t)$ near the change-points (shown by dashed vertical lines) are typically quite narrow, thus confirming a stepwise nature of the unknown rate function $\lambda(t)$. Fig. 6 (right axes) also shows the corresponding cumulative graphs simulated via Poisson processes with rates $\hat{\lambda}(t)$ (blue) and $\bar{\lambda}(t)$ (red), contrasted with the original cumulative plot (black) for threshold exceedances observed in the concentration data (cf. Fig. 1). As could be anticipated, the integrated (cumulative) estimates demonstrate a much better fit to the data. In particular, the cumulative plots for CO indicate that the discrepancy between $\hat{\lambda}(t)$ and $\bar{\lambda}(t)$ around the first change-point may be due to a gradual variation in the cumulative rate; this issue will be examined in Section 5.5 below.

*5.3. Verification of the Poisson change-point model*

There is a simple way to verify statistically the fitted Poisson change-point model (with a step rate function) using the estimated locations of the change-points but free from knowledge of the corresponding step heights. Recall that the hypothetical Poisson process generating the occurrence of the threshold exceedances (parameterised by a step rate function $\lambda(t)$) is homogeneous between the consecutive change-points. According to Section 2.1, this implies that, conditionally on a given number of observations, the temporal locations of exceedances should be distributed uniformly between the change-points. To verify this statistically from the data, one can apply various tests as described in Section 3.5.

Specifically, for $NO_2$ there are two step intervals separated by the estimated change-point $s_1 = 2490$ and containing $n_0 = 257$ and $n_1 = 133$ exceedances, respectively. The observed values of the test statistic $U_n$ defined in equation (9) were calculated as $U_{n_0} = -0.4781$ and $U_{n_1} = -0.6892$, with the approximate (normal) p-values 0.6326 and 0.4907, respectively. The Military Handbook test (see formula (10)) gave the values $\chi^2_{2n_0} = 496.3036$ and $\chi^2_{2n_1} = 237.5954$, with the corresponding p-values 0.2954 and 0.1058. Finally, the Kolmogorov–Smirnov test statistic $D_n$ (see formula (11)) specialised to $D_{n_0} = 0.0633$ and $D_{n_1} = 0.0888$, with the p-values 0.2447 and 0.2454, respectively. Therefore, the fitted model passes all the tests of uniformity with the majority of p-values at least 25% and just one about 11%. For the other two pollutants, NO and CO, the results were similar being consistent for different tests; for example, the majority of p-values for the Kolmogorov–Smirnov test were well above 20%, with just three somewhat smaller but still insignificant: 0.1356 (NO, interval $(s_2, s_3)$) and 0.1296, 0.0953 (CO, intervals $(s_2, s_3)$, $(s_3, s_4)$, respectively).

*5.4. Statistical assessment of alternative Poisson models*

Further verification of the fitted non-homogeneous Poisson model may be motivated by challenging the lack of identified change-points on some intervals. In particular, a bare-eye inspection of the cumulative plot for $NO_2$ may suggest, as was mentioned in Section 4.2, that there are a few more change-points after the first most obvious one near $t \approx 2500$, even though the MCMC estimation did not reveal any extra change-points there. We can test this statistically by comparing two competing models on the interval $[s_1, T] = [2490, 6206]$, namely the homogeneous model $M_0$ with a constant rate (as a null hypothesis) and the model $M_2$ with one additional change-point (as an alternative hypothesis). Applying the methods presented in Section 3.5, the observed value of the test statistic $\Delta_n$ given by equation (12) (with $n = n_1 = 133$) was computed as 2.8949, and formula (13) yields the approximate p-value 0.1457. The approach based on the Bayes log-factor $\beta_{02} = 2\log B_{02}$ (see equation (18)) gives the value $\beta_{02} = 7.0167$, which, according to Table 2, constitutes strong evidence in support of the model $M_0$.

The same calculations can also be carried out for the other two pollutants, NO and CO, confirming that there are no "extra" change-points in addition to those already identified by the MCMC estimator. A more interesting line of questioning the fitted Poisson model is to assess the main assumption about a *stepwise* variation of the rate function $\lambda(t)$, as opposed to a more gradual change, say according to a log-linear model $M_1$ (see Section 3.5). This is especially relevant to the data for NO and CO, for which the cumulative plots shown in Fig. 1 may look as if the rate decreases gradually rather then through abrupt jumps. Naturally, this leads to the problem of testing the model $M_1$ against the alternative $M_2$ (for simplicity, we consider only one hypothetical change-point at a time). Starting with the NO data and using the goodness-of-fit test for $M_1$ described in Section 3.5, from equation (14) we found the maximum likelihood estimate for the log-linear slope $\hat{\beta} = 0.000313$ and applied various known tests to check if the modified exceedance times ($t'_i$) (see equation (15)) are uniformly distributed. Specifically, from the asymptotically normal test with the statistic $U_n$ defined in equation (9), the p-value obtained was just slightly bigger than 0.025, suggesting that the hypothesis $M_1$ should be rejected. Furthermore, the Military Handbook test (equation (10)) and the Kolmogorov–Smirnov test (equation (11)) produced the p-values 0.0176 and 0.0147, respectively, signalling a statistically significant departure from the model $M_1$. This was complemented by calculating the Bayes log-factor $\beta_{21} = 2\log B_{21}$ (i.e., for the change-point model $M_2$ against $M_1$) using equations (17), (18) and (19), which yields $\beta_{21} = 7.9469$ and hence strongly rejects $M_1$ as recommended by Table 2.

Similar results were obtained for CO, confirming that the log-linear model $M_1$ is less suitable than the fitted model with a step rate function, even though the p-values in the uniformity testing were a little higher (about 0.05 to 0.06), reflecting a slightly more gradual variation of the rate function (e.g., manifested by a greater number of change-points). The model $M_1$ was also strongly rejected in favour of $M_2$ on the grounds of the Bayes log-factor $\beta_{21} = 8.5426$, as recommended by Table 2.

One can also use these methods in order to assess statistically the quality of the fitted step-rate Poisson model on various specific time intervals. For instance, it is interesting to investigate more closely the rate behaviour for carbon oxide CO on the first two inter-change-point intervals (i.e., from $s_0 = 0$ to $s_2 = 1867$) where there is a noticeable discrepancy between the estimated step rate function $\hat{\lambda}(t)$ and the posterior mean rate function $\bar{\lambda}(t)$ (see Fig. 5). For this part of the CO data, the hypothetical log-linear model $M_1$ (conjectured at the end of Section 5.2) can be assessed as described above. The tests of uniformity of the modified occurrences ($t'_i$) (see equation (15)), based on the estimated log-linear slope $\hat{\beta} = 0.000360$, give relatively small p-values as follows: 0.0704 ($U_n$- test, equation (9)), 0.0289 (Military Handbook test, equation (10)) and 0.0550 (Kolmogorov–Smirnov test, equation (11)), while the Bayes log-factor for the competing change-point model $M_2$ against $M_1$ equals $\beta_{21} = 6.2838$. Hence, there is significant statistical evidence against $M_1$, and so the change-point model $M_2$ is reinstated as the most suitable one.

### 5.5. *Validation of the fitted Poisson model*

To validate the employed MCMC sampler as an adequate posterior estimator of the unknown step rate function, and to illustrate the quality of the MCMC identification performance, we simulated data sets from non-homogeneous Poisson processes with the estimated step function $\hat{\lambda}(t)$ as a rate function and again applied the MCMC algorithm to see if there were any significant differences in the estimates. The validation results are graphically presented in Fig. 7 showing an excellent match of the posterior estimates for the simulated data sets (red) with those obtained earlier for the observed data (blue), both in terms of the rate functions and cumulative plots. Note that the red plots across Figs. 3–5, belonging to the same validation type, confirm further the consistency of the MCMC output.

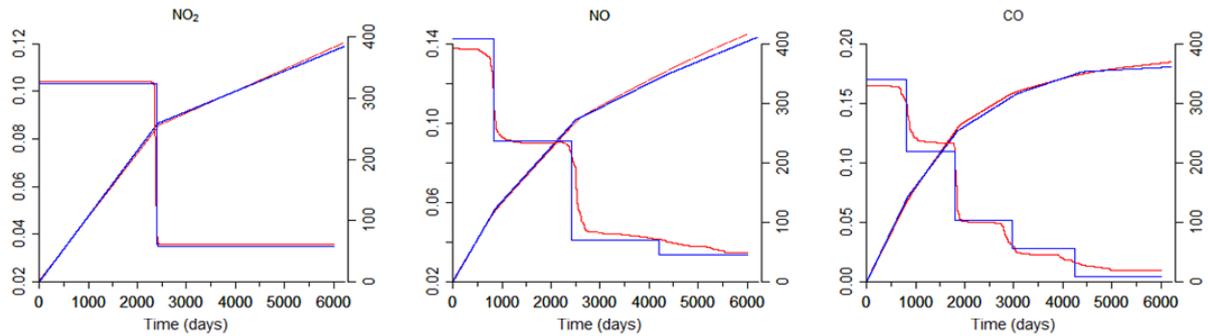

**Fig. 7.** Validation of the reversible jump MCMC performance in the step rate Poisson model estimation. The rate graphs (*left axes*) show the step rate function $\hat{\lambda}(t)$ (*blue*, same as in Fig. 6) estimated for the observed threshold exceedances from the MCMC posterior distribution, and the posterior mean rate $\bar{\lambda}(t)$ (*red*, different from Fig. 6) obtained for a data set simulated via a non-homogeneous Poisson process with rate $\hat{\lambda}(t)$. Cumulative plots (*right axes*) correspond to data sets simulated via non-homogeneous Poisson processes with rate $\hat{\lambda}(t)$ (*blue*) or $\bar{\lambda}(t)$ (*red*).

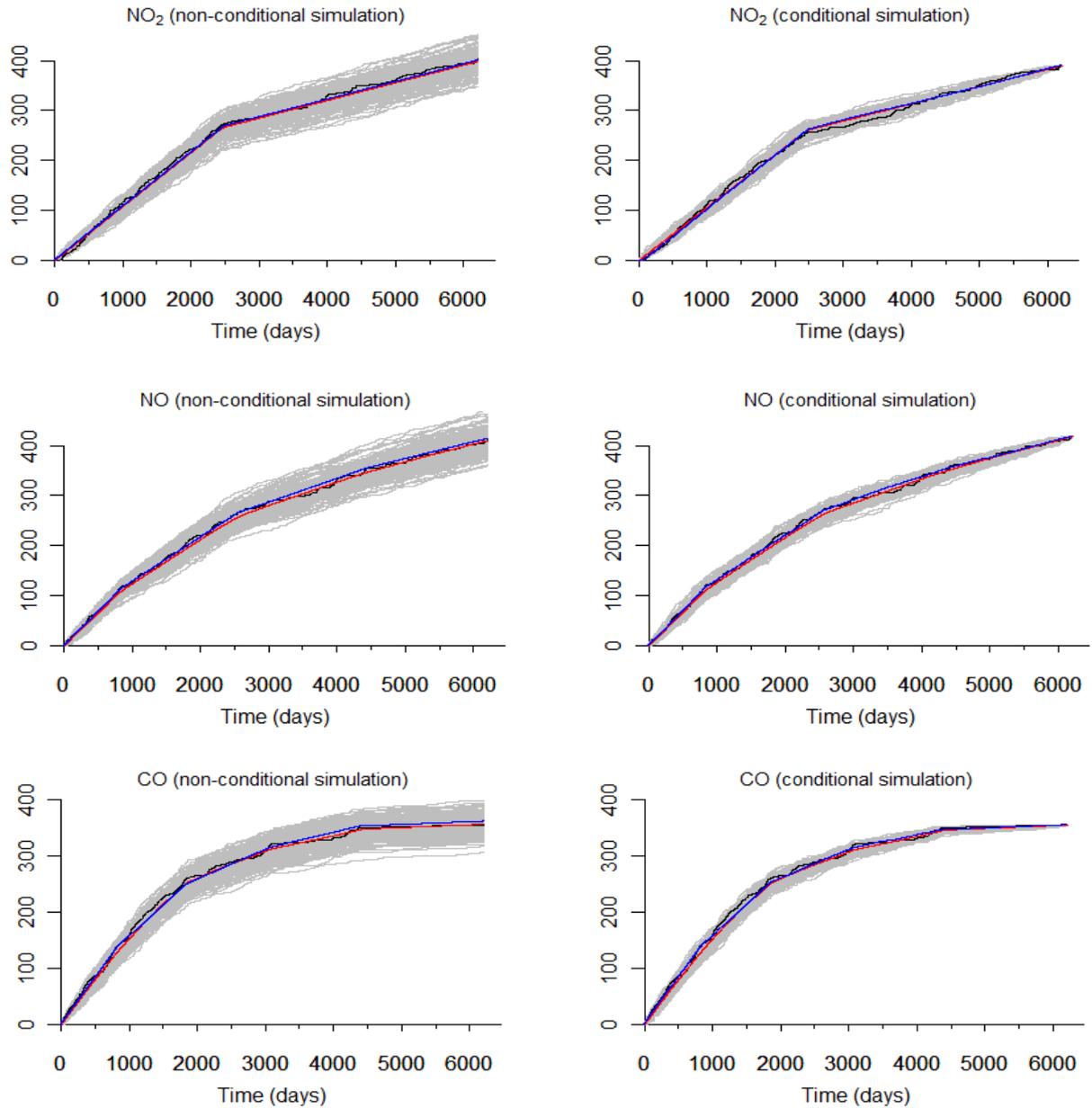

**Fig. 8.** Bayesian validation of the MCMC results using replicated counting processes. The graphs show cumulative plots for simulated data, non-conditional (*left*) and conditional (*right*) on the total number of observed threshold exeedances. *Black*: cumulative plots for the original exceedance data (same as in Fig. 6). *Grey*: 1000 replicated trajectories simulated each with a rate function independently sampled from the MCMC posterior distribution. *Red*: pointwise average of these trajectories. *Blue*: pointwise average of another 1000 simulated trajectories of a Poisson counting process with the step rate function $\hat{\lambda}(t)$ estimated from the posterior distribution.

An alternative (Bayesian) approach to the model validation is based on the so-called *posterior predictive simulation* (see Gilks et al., 1996) using several replicated data sets, each simulated via a step rate function drawn from the MCMC posterior distribution, and comparing them with the observed data. If the model is reasonably accurate, replicates should look similar to observations. We implement this idea by replicating the counting processes and comparing the corresponding cumulative plots (see Fig. 8). It is clear that, due to random fluctuations, the deviation of the replicated cumulative plots from the observed plot will increase in the course of time. Alternatively, noting that any Poisson process is time reversible (see Cox, 1981; Raftery, 1989), one could plot the cumulative counts in reverse time, whereby the deviations of the replicates from the observed cumulative plot would grow in the opposite direction. Combining these two approaches, the accuracy of replicated cumulative plots may be significantly improved by simulating *conditional* Poisson processes subject to the constraint that the total number of Poisson occurrences is conditioned to be the same as that actually observed in the real data. Both non-conditionally and conditionally replicated cumulative counts are presented in Fig. 8, showing a fairly good fit with the observed (declustered) cumulative plot and also consistent with the sample mean plot for data sets simulated with a posterior (step) estimate $\hat{\lambda}(t)$.

## 6. Discussion

### 6.1. General comments on the model and results

Statistical analysis of observed pollutant concentrations developed in this paper is essentially based on the extreme value approach. The data at hand involved the maximum daily concentration levels as raw measurements and, more importantly, high threshold exceedances which were utilised for statistical inference. We focused on the corresponding point process of exceedance occurrences in time, modelled via a non-homogeneous Poisson process with a step rate function, whereas the actual values of exceedances were not taken into consideration. An advantage of this simplified model (even though still complex enough being set out in a parametric space of varying dimension) is that it may be expected to be reasonably robust, that is, rather insensitive to deviations from the model assumptions, and hence to have a more flexible potential for real data applications. In fact, the algorithmic tractability of the model makes it suitable for an efficient MCMC multi-parametric estimation. As has been demonstrated in this work, the reversible jump MCMC algorithm applied to the multiple change-point Poisson model provides a reliable identification of the unknown change-points in the rate function, including their number, location and the corresponding step heights.

A global downward trend in the behaviour of the rate functions for each pollutant (see Fig. 6) indicates a general improvement in the air quality, arguably promoted by environmental actions (NAEI, 2010). Such a trend could also be attributed to the ongoing improvement of car engines, in particular, using the three-way catalytic converters as standard (NAEI, 2010), and to a gradual change in drivers' attitudes and habits (Glaister, 2002), stimulated by the growing fuel prices and the government's pressure on "going green", for instance by choosing newer and smaller cars with a less powerful engine or by switching from petrol to diesel, primarily as being more economical but also emitting less CO, which is reflected in a lower road tax (also known as Vehicle Excise Duty). Clearly, such changes develop gradually, so the change-point model with a step rate function is not suitable for capturing this kind of dynamics. Even though the model $M_1$ with a log-linear rate function (Section 3.5) was rejected at every test, both globally and on selected specific intervals (see Section 5.4), it would be interesting to combine the two models by considering a more general class of rate functions (cf. equation (5) and the definition of model $M_1$ in Section 3.5):

$$\lambda(t) = e^{-\beta t}\left(\sum_{j=0}^{k} h_j \mathbf{1}\{s_j \leq t < s_{j+1}\}\right), \qquad 0 = s_0 < s_1 < \cdots < s_k < s_{k+1} = T. \tag{24}$$

The model (24) can be reduced to a multiple change-point model (5) by filtering out the log-linear term as part of the deseasonalisation procedure described in Section 3.4, whereby the regression model (7) will now include the log-linear term suggested by equation (24), leading to an additional unknown parameter $\beta$:

$$\log \widetilde{X}(t) = \log X(t) + \beta t - a\cos\omega t - b\sin\omega t - c. \tag{25}$$

Figure 9 illustrates the results of fitting the combined log-linear/periodic trend (25) to the raw data; note a better fit of the new trend (shown as a red oscillating curve) as compared to a purely periodic trend (cf. Fig. 2).

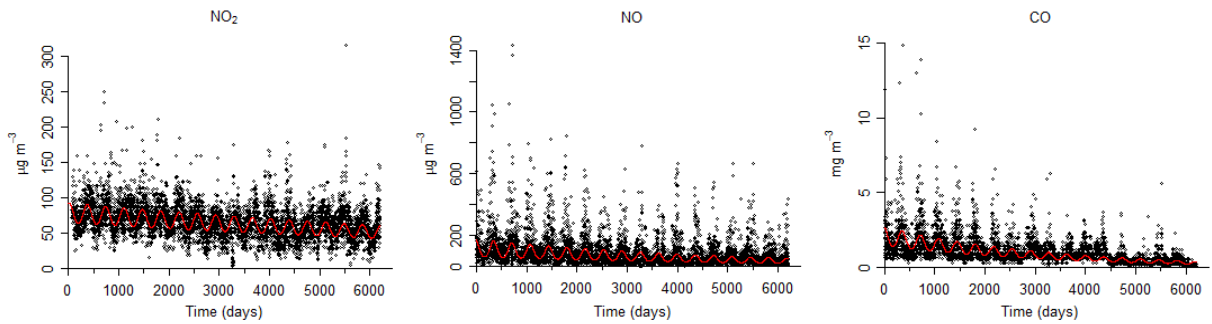

**Fig. 9.** Combined log-linear/periodic trend of the form (25) (*red*) fitted to the the log-transformed raw data (*black*).

However, after several rounds of experimentation (not reported here) it was decided not to include the log-linear part in our model, because it would tend to overshadow the identification of change-points in the step part of the rate function. The reason for such an effect is not quite clear, one possible explanation being that taking out the log-linear trend may mask the prominence of high exceedances, especially towards the beginning of observations. Thus, the problem of an efficient MCMC change-point identification in the combined model (24) remains to be investigated more thoroughly, but it is the more remarkable that even a simplified, purely stepwise rate model considered in this paper is capable of producing convincing results, as has been demonstrated in Sections 5.2–5.5.

To conclude our general comments, let us summarise some points of novelty achieved in the present paper as compared to the previous literature on the Poisson-based air quality modelling.

- We have adapted a reversible jump MCMC Metropolis–Hastings sampler, proposed by Green (1995), in the context of high threshold exceedances for pollutant concentration data. A successful implementation of this technique has been possible thanks to a relatively low complexity of a multiple change-point Poisson model and, on the other hand, due to a careful preprocessing of the data (Sections 3.1–3.4).

- Unlike many other existing approaches, we have assumed a very simple non-homogeneous Poisson model with a step rate function, instead of working with more complicated parametric classes (cf. Shively, 1991; Achcar et al., 2008, 2009a). Due to a reduced number of the model parameters, the reversible jump MCMC estimator has been capable of producing good results even on relatively small data sets of exceedances. Note that no prior restrictions were imposed on the unknown step rate function including its values (e.g., no monotonicity was postulated) and the number of change-points. Achcar et al. (2009a, 2009b) considered a more complicated change-point Poisson model with a certain parametric rate function, but they were unable to handle more than three change-points due to a rapidly growing computational complexity of their hierarchical MCMC algorithm.

- A simple and efficient method for imputing missing data has been used (Section 3.2), thus improving on an earlier approach by Smith and Shively (1995). This may be contrasted with a series of papers by Achcar et al. (2008, 2009a, 2009b) analysing the Mexico City ozone data (www.sma.df.gob.mx/simat/) that contained a significant proportion of missing values.

- Deseasonalisation (i.e., elimination of yearly cycles in the data) has been carried out prior to identification of the change-points (Sections 3.3), thus dramatically enhancing the MCMC performance as confirmed by our simulation studies, even though the weekly oscillations present in the data, as well as a plausible log-linear trend, have not been taken into account.

- A thorough procedure of declusterisation has been applied in order to remove inter-dependence among the threshold exceedance occurrences (Sections 3.4, 5.1). Statistical testing of independence has been extensively used as a diagnostic tool to assess the quality of declusterisation and ensure the validity of the non-homogeneous Poisson model, including cross-validation via running MCMC iteratively. Note that the important issue of declusterisation is not always addressed (or even mentioned) in the applied literature in the air quality context, the Poisson model often being just postulated (cf. Achcar et al., 2008, 2009a).

- The results of MCMC model identification and estimation (Section 5.2) have been examined and assessed (Sections 5.3, 5.4) by applying a variety of statistical tests for goodness-of-fit and for hypothesis testing to select the best fitting model. The tests included both classical and Bayesian tools (Section 3.5).

- The performance of the reversible jump MCMC estimator has been validated using simulated data sets, including multiple replications from the posterior distribution (Section 5.5). In particular, the new conditional version of simulated cumulative plots (subject to the known number of events) has been proposed, giving a much improved fit to the observed data. The results of validation have provided us with useful information about how MCMC works in practice, which will be helpful in adjusting the MCMC technique to more complex models.

*6.2. Mapping and interpretation*

In this work, we have made a preliminary attempt to consider the "mapping" of our results, that is, to relate the identified change-points presented in Table 3 with available accounts and records of the real transport-related circumstances and events over the period under study (i.e., from 4 January 1993 to 31 December 2009), both from official sources and the media. A few examples of that kind (with or without such a relation) are given below starting with possible local effects (6.2.1–6.2.5) followed by global ones (i.e., national or European, 6.2.6–6.2.7

It should be stressed that, due to the computational complexity of the model and the distribution-oriented nature of Bayesian inference employed, any specific conclusions of the MCMC-based analysis (e.g., about the number and locations of the change-points) must be treated with due caution and consideration, and certainly not taken as an ultimate diagnosis. There is a need for an additional cross-examination and verification of these findings in co-operation with experts from other subject areas, including policy makers as well as environmental and transport modellers and practitioners, especially those who have more information about past environmental actions. Special care is required when relating the conclusions of the threshold exceedance analysis with official statistics (see, e.g., NAEI, 2010; Faulkner and Russell, 2010) usually based on averaged observations rather than on extreme values. This underpins possible significant discrepancies in the interpretation, since extreme value statistics may appear more sensitive to traffic pollution changes than the typical, bulk values (Smith, 2004). For instance, conventional summary statistics such as the median may not be sensitive to narrow high spikes in the concentration data, while these will be picked up by our model as threshold exceedances.

*6.2.1. High Occupancy Vehicle (HOV) lane.* According to Table 3, the $1^{st}$ change-point for $NO_2$ (29 October 1999) and the $2^{nd}$ change-point for NO (11 November 1999) are very close to each other. Arguably, both may be attributable to the possible impact of the High Occupancy Vehicle (HOV) lane scheme, implemented in Leeds as part of the European ICARO research project and designed to give priority to vehicles with more than one commuter, thus aiming to increase car occupancy (HOVL, 2004). The HOV lane in Leeds was opened on 11 May 1998 under an experimental Traffic Regulation Order (and made permanent from 8 November 1999) on the A647 Stanningley Road and Stanningley By-Pass, which form the principal radial route to the west of the city centre. According to the official assessment (HOVL, 2004), there was a relatively small improvement of air quality on the A647, claimed to be attribut-able "to reduced vehicle emissions rather than to the impact of the HOV lane". The document did not provide any further argumentation for such a conclusion, but our results suggest that the HOV lane may have had a more prominent role in the air quality improvement. Even though our study is based on the data (AQA, 2010) collected about 5 miles east of the HOV lane, it would not be completely unrealistic to assume that the reduced congestion on the A647 road could also affect the traffic in the city centre. In addition, due to predominantly westerly winds in North East England (MET-O, 2010), an improved air quality near Stanningley Road might positively contribute to the city centre. Of course, this is no more than a crude preliminary assessment that needs to be elaborated on the basis of detailed (daily) information about the wind, which highlights the importance of accounting for meteorological covariates in the pollution analysis.

*6.2.2. East Leeds Link Road.* As part of its Air Quality Action Plan (AQAP, 2004), Leeds City Council initiated a number of projects aimed to improve the highway network, including the East Leeds Link Road (ELLR) designed to reduce through-traffic in the city centre (ELLR, 2010). The works on the ELLR construction started in November 2006, and the new 4 km dual carriageway link between junction 45 of the M1 and the Leeds Inner Ring Road was opened on 10 February 2009. The new road was expected to provide "indirect" improvements in the air quality, with low to moderate impact (AQAP, 2004), but no real data to this effect have been made available as yet. Our model has not detected any change-points (for either of the pollutants) around February 2009 or after, but using more data beyond December 2009 might make it different. Note that due to a general downward trend in both average raw concentrations (see Fig. 2) and threshold exceedances (see Fig. 6), keeping the threshold fixed (say, at a 90% empirical level for the entire data set, as in the present work) is likely to cause the loss of diagnostic power of the MCMC identification method in the course of time. In order to avoid such "ageing" and to maintain the sensitivity of the MCMC sampler, it seems necessary to update (perhaps dynamically) the threshold level to be used for more recent data.

*6.2.3. Leeds Free City Bus.* This zero-fare service (3.4 miles long loop) in the Leeds city centre began on 30 January 2006. It was the first scheme of this format in West Yorkshire. It is estimated that the Free City Bus has reduced the year-on-year numbers of cars entering the city centre (LCCA, 2010, p. 29) and hence is contributing towards meeting a target set by the current West Yorkshire Local Transport Plan of 10% reduction by 2011 in nitrogen dioxide $NO_2$ in the Leeds air quality management areas (LCCA, 2010, p. 31). Again, the impact of the scheme was not strong enough to be picked up by our MCMC model, perhaps being masked by the general gradual downward trend.

*6.2.4. "Christmas 2004" change-points.* Note that there are two different change-points in Table 3 that are quite close to each other, namely for NO (day 4350, 1 December 2004) and CO (day 4388, 8 January 2005), with overlapping 50%-credible intervals. It remains to investigate further if this can be associated with any known environmental or transport-related policies introduced in Leeds around that time (but see Section 6.2.6 below). Remarkably, both change-points are close to the season of Christmas and New Year holidays, when the traffic drops dramatically in all categories, but especially for trucks and vans (light commercial vehicles, LCV) and lorries (large goods vehicles, LGV). It is unclear if this alone has affected the threshold exceedance data in this particular year so much as to lead to a statistically significant change-point for both NO and CO (but not for $NO_2$). Incidentally, note that there is a very strong drop in the raw time series for $NO_2$ localised approximately around day 3270 (see Fig. 2), which falls on 16 December 2001, being within a Christmas season as well; however, no change-points were detected for $NO_2$ anywhere near.

6.2.5. *Construction works.* There have been ongoing redevelopment and construction works near the AURN monitoring station in the Leeds city centre, along with the associated transportation and engineering activities. The project started in mid-January 2005 and continued until end of June 2009 when the construction site was dismantled and landscaped (AURN, 2010). There have been some concerns that the site activities might be responsible for sporadic elevated pollution levels in the area, so it is interesting to investigate if our results have been sensitive to these hypothetical effects. By inspection of Table 3, we note that the latest identified change-point across all the three pollutants under study was on 8 December 2005 (for CO), with the 50% credible interval from 17 November 2004 to 1 April 2005 covering the commencement date of the aforementioned construction project. However, a causal association seems unlikely, because this change-point resulted in a decrease of the estimated rate function, from $h_4 = 0.0228$ to $h_5 = 0.0027$ (see Table 4), and so the effect was in fact opposite to what one might suspect. Hence, one can argue that there was no statistically significant impact of the construction works on the threshold exceedance data, suggesting that the related contribution to the air pollution in $NO_2$, NO and CO was relatively mild as compared to a presumably dominant input from intensive traffic on the nearby roads (AURN, 2010). On the other hand, it should not be ruled out that contributions to other pollutants might have been more prominent (e.g., particulate matter due to dust) and hence could be picked up by our methods; indeed, $PM_{10}$ concentration data (not reported here) reveal substantially elevated levels from April to late December 2005. This issue, which underscores the importance of a multivariate approach in air quality statistical inference, will be examined in more detail in our future work.

6.2.6. *Fuel crisis 2000.* The UK "fuel crisis", lasting about one week in September 2000, was caused by the announced increase of the fuel prices and subsequent public protests including blockade of oil facilities, which led to considerable disruption of fuel deliveries and hence to significant reductions in traffic flows (Hathaway, 2001). According to the Leeds City Council assessment (AQAP, 2004, p. 10), this resulted in a dramatic improvement of the air quality in the city (e.g., on average by 30–40% for $NO_2$). Our model has not been able to pick up any noticeable change anywhere near this point in time, which is not surprising in view of a very short-lived impact of this episode.

6.2.7. *European emission standards.* These standards, determining acceptable exhaust emission limits for new vehicles sold within the European Union, are defined through staged European Union directives introducing increasingly stringent limits on vehicle emissions (EUR, 2003). The regulations are set out separately for different types of vehicles (e.g., passenger cars, LCV, LGV, buses, etc.) and fuel (diesel or petrol). For example, the inception dates of the new emission standards for passenger cars (both diesel and petrol) were as follows: July 1992 (Euro 1), January 1996 (Euro 2), January 2000 (Euro 3), January 2005 (Euro 4) and September 2009 (Euro 5). By comparison with Table 3, we note that some of these dates can be related to the identified change-points. In particular, the pair mentioned in Section 6.2.1 (i.e., 1st change-point for $NO_2$, 29 October 1999, and 2nd change point for NO, 11 November 1999) are close to the starting date of Euro 3 (covered by the 50% credible intervals), while the Christmas 2004 change-points (i.e., 1 December 2004 for NO and 8 January 2005 for CO, see Section 6.2.4) may be linked with the starting date of Euro 4. Furthermore, the earlier change-point for NO (18 May 1995) and the nearby change-point for CO (5 April 1995) may be attributed (even though with somewhat lower credibility) to Euro 2 which significantly tightened up the CO standard for diesel cars from 2.72 to 1.0 g/km and also reduced the $NO_x$ permitted levels by 25–45% (EUR, 2003, p. 155).

6.3. Future work

Let us outline here a few possible directions for future research to extend and enhance the MCMC techniques developed in the present work.

- It may be interesting to experiment with Green's reversible jump MCMC sampler by modifying some of its settings. For instance, it may be useful to be able to tune in the algorithm by varying its birth-death probabilities (see equation (22)) by making the random walk more aggressive (to enhance the exploration) or more conservative.

- There is a need to overcome the unrealistic assumption of a pure-step structure of the rate function, for example by combining a step rate with a log-linear trend as in equation (24). This approach would allow one to capture continuously decreasing trends between change-points as a manifestation of ongoing improvements in car engines and fuels.

- Possible correlations in the exceedance time series between the change-points can be modelled using a *renewal process* (Cox and Lewis, 1966), whereby the random waiting times $\tau_i$ between successive occurrences may have a distribution other than the exponential one (see Section 2.1). One suitable class of renewal processes is specified by choosing the Weibull distribution $P(\tau_i > t) = \exp(-\lambda t^\beta)$, which allows one to model "ageing" ($\beta > 1$) or "rejuvenation" ($\beta < 1$) of the waiting times. In order to adapt the reversible jump MCMC sampler to the renewal case, formulas for updates and their accep-tance probabilities must be modified to ensure the reversibility of the corresponding Markov chain (cf. Section 4.3).

- Identification of the change-points obtained in the present study should be cross-examined by running the same MCMC sampling algorithm on different data sets independently collected at the same or nearby location. Such an opportunity arises in Leeds due to a fixed roadside laboratory station located on A65, Kirkstall Road (about 2 miles away from the city centre). The laboratory also houses a traffic monitoring system and a meteorological station, which provides additional useful information that may facilitate the interpretation of results.

- The simplified univariate Poisson model used in the present work may be extended by analysing both the times and magnitudes of threshold exceedances, which can be approximated by a two-dimensional (non-homogeneous) Possion process (Leadbetter et al., 1983). Such an approach, pioneered by Smith and Shively (1995) for certain parametric families of the rate functions, can be developed to include change-points describing possible abrupt changes in the observed time series. The emphasis of this generalisation lies with building and tuning a suitable reversible jump MCMC algorithm.

- The study in this paper has been limited to three pollutants, $NO_2$, NO and CO, in order to test the methodology and not to obscure the analysis with too much data. At the next stage, it is of interest to apply our methods to the concentration data for other pollutants such as carbon dioxide $CO_2$, ozone $O_3$ and particulate matter $PM_{10}$, using the same air quality data base (i.e., AQA, 2010) and the same time frame (cf. Section 6.2.5).

- Many air pollutants are correlated due to photochemical and other reactions (cf. Section 3.3). Therefore, to gain more information from the individual time series for separate pollutants, it is important to develop a multivariate MCMC algorithm efficiently processing several pollutants at a time.

## 7. Conclusions

In this paper, we have used a non-homogeneous Poisson model to fit the observed series of threshold exceedance occurrences in time, extracted from the daily pollution concentration data obtained in the City of Leeds, UK, in 1993–2009. For simplicity, the time dependence of the unknown Poisson rate was assumed to be a step (i.e., piecewise constant) function, with an arbitrary number of possible change-points and no restrictions on its values (in particular, with no *a priori* requirement of monotonicity). The principal aim of the modelling was to identify statistically significant change-points, which might indicate notable changes in the emission-driven air pollution levels. Once the statistical identification is carried out and validated (e.g., by simulations), it is important to seek a meaningful interpretation of the results, bearing in mind updates in the environmental policies by local and central authorities as one possible source of the emerging change-points.

Certain care is needed in practical applications of the Poisson model, since additional "false" change-points may be induced in the data by meteorological conditions, seasonal effects, missing values and cluster correlations in threshold exceedances. We have addressed these issues by an appropriate preprocessing of the data (such as deseasonalisation and declusterisation), supported by the graphical and statistical assessment of the results.

Statistical estimation of the unknown rate function was carried out using Bayesian posterior samples obtained by adapting a reversible jump Markov chain Monte Carlo (MCMC) simulation algorithm first proposed by Green (1995) to handle models of unspecified dimension. The results of our work have demonstrated the computational and statistical efficiency of the MCMC method.

The application of non-homogeneous Poisson processes with multiple change-points may provide a suitable modelling framework in the air quality management, which can be used to analyse pattern changes in the violation of air quality standards by various pollutants. In particular, these methods may be instrumental in obtaining important feedback about environmental actions and assessing their impact on the spatiotemporal dynamical patterns of potentially hazardous pollutants.

## Acknowledgements

J. Gyarmati-Szabó was supported by an EPSRC Doctoral Training Grant, by the Strategic Fund of the Institute for Transport Studies (ITS), University of Leeds, and by a Postgraduate Research Scholarship of the School of Mathematics, University of Leeds. L.V. Bogachev was partially supported by a Leverhulme Research Fellowship.